\documentstyle[11pt,twoside,fleqn,espcrc1,epsfig]{article}

\newcommand{\AmS}{{\protect\the\textfont2
  A\kern-.1667em\lower.5ex\hbox{M}\kern-.125emS}}

\title{Generalized polarizabilities and the chiral structure of the nucleon}

\author{Thomas R.\ Hemmert,\address{TRIUMF, Theory Group, 4004 Wesbrook Mall,
Vancouver, British Columbia, Canada V6T 2A3}
Barry R. Holstein,\address{Department of Physics and Astronomy,
University of Massachusetts, Amherst, Massachusetts 01003}
Germar Kn{\"o}chlein,$^{\footnotesize c}$ and 
Stefan Scherer\address{Institut f\"ur 
Kernphysik, Johannes 
Gutenberg-Universit\"at, D-55099 Mainz, Germany}
\footnote{Talk given at
the 15th International Conference on Few Body Problems in Physics,
Groningen, The Netherlands, 22-26 July 1997}
}

\begin{document}
% typeset front matter
\maketitle

\begin{abstract}
   We discuss the virtual Compton scattering reaction $e^-p\to e^-p\gamma$ 
at low energies.
   We present results for the generalized polarizabilities of the nucleon
obtained in heavy baryon chiral perturbation theory at $O(p^3)$. 
\end{abstract}

\section{INTRODUCTION}
   At energies below the pion-production threshold, the amplitude for Compton
scattering off the nucleon may be expanded in a Taylor series in the momentum
of the photon.
   The famous low-energy theorem (LET) of Low \cite{Low_1954} and  
Gell-Mann and Goldberger \cite{GellMann_1954} makes a model-independent
prediction for the expansion up to and including linear terms.
   The Taylor series coefficients are given in terms of the 
charge, mass, and magnetic moment of the target, i.e.\ properties of the 
nucleon which can entirely be determined from different experiments.
   The derivation of this theorem is based on gauge invariance, Lorentz 
covariance, crossing symmetry, and the discrete symmetries.
   The description of second-order terms requires two new structure
constants specific to the real Compton scattering amplitude, 
the electric and magnetic polarizabilities $\bar{\alpha}_E$ and 
$\bar{\beta}_M$ (for an overview see, e.g., \cite{Holstein_1992}).

   The investigation of virtual Compton scattering (VCS) as 
tested in, e.g., the reaction $e^-p \to e^-p \gamma$, has recently attracted 
a lot of interest.
   Even though the experiments \cite{VCSexp} will be considerably more 
complicated than for real Compton scattering, there is the prospect 
of obtaining completely new information about the structure of the nucleon
which cannot be obtained from any other experiment.
   Like in real Compton scattering, the model-independent
properties of the low-energy VCS amplitude have been identified in
\cite{Guichon_1995,Scherer_1996}. 
   In \cite{Guichon_1995} the model-dependent part beyond the LET was analyzed
in terms of a multipole expansion.  
   Keeping only terms linear in the energy of the final photon, 
the corresponding amplitude was parametrized in terms of 
ten so-called generalized polarizabilities (GPs) which are functions
of the three-momentum transfer of the initial virtual photon.
   The number of independent GPs reduces to six, if 
charge-conjugation invariance is imposed 
\cite{Drechsel_1997a,Drechsel_1997b}.

\section{KINEMATICS AND LET}
   Omitting the Bethe-Heitler contribution, the invariant amplitude for VCS 
reads
\begin{equation}
\label{mvcs}
{\cal M}_{VCS}=-ie^2\epsilon_\mu \epsilon_\nu'^\ast M^{\mu\nu}
=-ie^2\epsilon_\mu M^\mu
=ie^2\left(\vec{\epsilon}_T\cdot\vec{M}_T
+\frac{q^2}{\omega^2}\epsilon_z M_z\right),
\end{equation}
where $\epsilon_\mu=e\bar{u}\gamma_\mu u/q^2$ 
is the polarization vector of the virtual photon
($e>0,e^2/4\pi \approx 1/137$), and where use of 
current conservation has been made.
   In the center-of-mass system, using the Coulomb gauge for the final
real photon, the transverse (longitudinal) part of ${\cal M}_{VCS}$ 
can be expressed in terms of eight (four) independent structures
\cite{Scherer_1996,Hemmert_1997a}, 
\begin{equation}
\vec{\epsilon}_T\cdot \vec{M}_T=\vec{\epsilon}\,'^\ast \cdot 
\vec{\epsilon}_T A_1 + \cdots,\quad
M_z=\vec{\epsilon}\,'^\ast \cdot \hat{q} A_9 + \cdots,
\end{equation}
   where the functions $A_i$ depend on three kinematical variables,
$\bar{q}=|\vec{q}|$, $\omega'=|\vec{q}\,'|$, $z=\hat{q}\cdot\hat{q}\,'$.

   Extending the method of Gell-Mann and Goldberger \cite{GellMann_1954} to
VCS, model-independent predictions for the functions $A_i$
%, based on Lorentz covariance, gauge invariance, crossing symmetry, and 
%the discrete symmetries 
were obtained in \cite{Scherer_1996}.
   For example, the result for $A_1$ up to second order in $\bar{q}$
and $\omega'$ is
\begin{eqnarray}
\label{a1}
A_1&=&-\frac{1}{M}+\frac{z}{M^2}\bar{q}
-\left(\frac{1}{8M^3}+\frac{r^2_E}{6M}-\frac{\kappa}{4M^3}
-\frac{4\pi\bar{\alpha}_E}{e^2}\right)\omega'^2\nonumber\\
&&+\left(\frac{1}{8M^3}+\frac{r^2_E}{6M}-\frac{z^2}{M^3}
+\frac{(1+\kappa)\kappa}{4M^3}\right)\bar{q}^2.
\end{eqnarray}
   To this order, all $A_i$ can be expressed in terms of 
%known quantities, namely, 
$M$, $\kappa$, $G_E$, $G_M$, $r^2_E$, $\bar{\alpha}_E$, 
and $\bar{\beta}_M$.

\section{GENERALIZED POLARIZABILITIES}

   For the purpose of analyzing the model-dependent terms 
specific to VCS, the invariant amplitude is split into a
pole piece ${\cal M}_A$ and a residual part ${\cal M}_B$.  
   The s- and u-channel pole diagrams are calculated using electromagnetic 
vertices of the form
\begin{equation} \label{f1f2vertex}
\Gamma^\mu(p',p)=\gamma^\mu F_1(q^2)+i\frac{\sigma^{\mu\nu}q_\nu}{2M} F_2(q^2),
\,q=p'-p, \end{equation}
where $F_1$ and $F_2$ are the Dirac and Pauli form factors, 
respectively.
   The corresponding amplitude ${\cal M}^{\gamma^\ast\gamma}_A$ contains all 
irregular terms as $q\to 0$ or $q'\to 0$ and is separately gauge invariant.  
 
   The generalized polarizabilities in VCS \cite{Guichon_1995} result from 
an analysis of ${\cal M}_{B}^{\gamma^{\ast} \gamma}$ in terms of 
electromagnetic multipoles $H^{(\rho' L', \rho L)S}(\omega' , \bar{q})$,  
where $\rho \, (\rho')$ denotes the type of the initial 
(final) photon ($\rho = 0$: charge, C; $\rho = 1$: magnetic, M;
$\rho = 2$: electric, E).  
   The initial (final) orbital angular momentum is denoted by 
$L \, (L')$, and $S$ distinguishes between non-spin-flip $(S = 0)$ and 
spin-flip $(S = 1)$ transitions.
   According to the LET for VCS, ${\cal M}^{\gamma^\ast\gamma}_B$ is at 
least linear in the energy of the real photon.
   A restriction to the lowest-order, i.e.\ linear terms in $\omega'$ leads 
to only electric and magnetic dipole radiation in the final state.
   Parity and angular-momentum selection rules (see Table 
\ref{tab:mulan}) then allow for 3 scalar 
multipoles $(S = 0)$ and 7 vector multipoles $(S = 1)$. 
\begin{table}[hbt]
% -----------------------------------------------------
% adapted from TeX book, p. 241
\newlength{\digitwidth} \settowidth{\digitwidth}{\rm 0}
\catcode`?=\active \def?{\kern\digitwidth}
% -----------------------------------------------------
\caption{Multipolarities of initial and final states}
\label{tab:mulan}
\begin{tabular}
%{\textwidth}{@{}l@{\extracolsep{\fill}}
{|c|c|c|}
\hline
$J^P$&final real photon&initial virtual photon\\
\hline
$\frac{1}{2}^-$&E1&C1,E1\\
\hline
$\frac{3}{2}^-$&E1&C1,E1,M2\\
\hline
\hline
$\frac{1}{2}^+$&M1&C0,M1\\
\hline
$\frac{3}{2}^+$&M1&C2,E2,M1\\
\hline
\end{tabular}
\end{table}
   The 10 GPs, $P^{(01,01)0}$, ...,
${\hat{P}}^{(11,2)1}\,$,
are functions of $\bar{q}^2$ and are related to the 
corresponding multipole amplitudes as 
\begin{eqnarray}
\label{gl3_1}
P^{(\rho' L' , \rho L)S} (\bar{q}^2)
& = &
\left[ \frac{1}{\omega'^{L} \bar{q}^{L}}
H^{(\rho' L' , \rho L)S} (\omega' , \bar{q}) \right]_{\omega' = 0} \, ,
\\ \label{gl3_1b}
\hat{P}^{(\rho' L' , L)S} (\bar{q}^2) 
& 
= 
&
\left[ \frac{1}{\omega'^{L} \bar{q}^{L+1}}
\hat{H}^{(\rho' L' , L)S} (\omega' , \bar{q}) \right]_{\omega' = 0}
\,,
\end{eqnarray}
   where mixed-type polarizabilities, 
${\hat{P}}^{(\rho' L' , L)S} (\bar{q}^2)$, have been introduced which are 
neither purely electric nor purely Coulomb type (see \cite{Guichon_1995}
for details). 
   Only six of the above ten GPs are independent, if charge-conjugation 
symmetry is imposed \cite{Drechsel_1997a,Drechsel_1997b}.

\section{RESULTS IN CHIRAL PERTURBATION THEORY}

   The calculation of the generalized polarizabilities is performed within 
the heavy-baryon formulation of chiral perturbation theory (HBChPT) 
\cite{HBChPT} to third order in the external momenta.
   At $O(p^3)$, contributions to the GPs are generated by nine one-loop 
diagrams and the $\pi^0$-exchange $t$-channel pole graph (see
\cite{Hemmert_1997a}).   
   For the loop diagrams only the leading-order Lagrangians are
required \cite{HBChPT},
\begin{equation}
\label{lpin1pipi2}
\widehat{\cal L}_{\pi N}^{(1)}=\bar{N}_v (i v \cdot D + g_A S \cdot u) N_v, 
\quad 
{\cal{L}}_{\pi \pi}^{(2)} = 
\frac{F_{\pi}^2}{4} 
\mbox{Tr}\left[ \nabla_{\mu} U (\nabla^{\mu} U)^{\dagger}\right],
\end{equation}
   where $N_v$ represents a non-relativistic nucleon field, and 
$U = {\mathrm{exp}}(i \vec \tau \cdot \vec \pi/F_{\pi})$ contains the pion 
field. 
   The covariant derivatives $\nabla_{\mu}U$ and $D_{\mu} N_v$ include
the coupling to the electromagnetic field, 
   and $u_\mu$ contains in addition the derivative coupling of a pion.
   In the heavy-baryon formulation the spin matrix is given by
$S^{\mu} = i \gamma_5 \sigma^{\mu\nu} v_{\nu}$, where $v^\mu$ is a
four-vector satisfying $v^2=1, v_0\ge1$ \cite{HBChPT}.
   Finally, for the $\pi^0$-exchange diagram we require in addition to
Eq.\ (\ref{lpin1pipi2}) the $\pi^0\gamma\gamma^\ast$ vertex 
provided by the Wess-Zumino-Witten Lagrangian,
\begin{equation}
\label{wzwpi0}
{\cal{L}}_{\gamma\gamma\pi^0}^{(WZW)} =  -\frac{e^2}{32\pi^2 F_\pi} \;
\epsilon^{\mu\nu\alpha\beta} F_{\mu\nu} F_{\alpha\beta} \pi^0 \,,
\end{equation}
   where $\epsilon_{0123}=1$ and $F_{\mu\nu}$ is the electromagnetic field
strength tensor.

\begin{figure}[ht]
\begin{minipage}[ht]{160mm}
%\framebox[79mm]{\rule[-26mm]{0mm}{52mm}}
\caption{GPs of the proton as a function of $\bar{q}^2$. 
The dashed line is the contribution from pion-nucleon loops, 
the dotted one from the $\pi^0$
exchange graph and the dash-dotted line the sum of both.
\label{fig:GPs}}
%\begin{center}
\epsfig{file=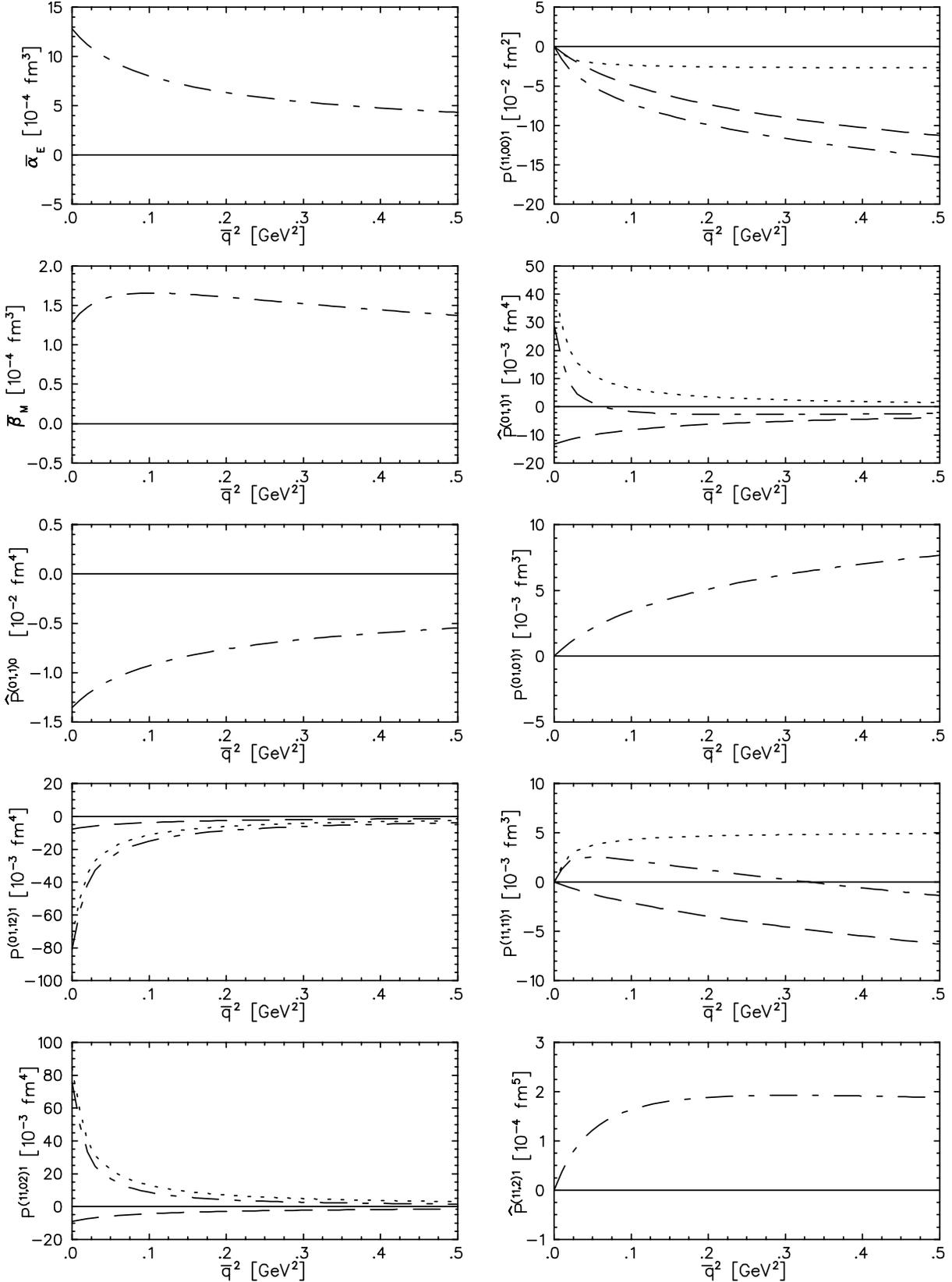,width=16cm}
%\end{center}
\end{minipage}
\end{figure}

   At $O(p^3)$, the LET of VCS is reproduced by the 
tree-level diagrams obtained from Eq.\ (\ref{lpin1pipi2}) and the relevant 
part of the second- and third-order Lagrangian 
\cite{Ecker_1996},
\begin{eqnarray}
\label{lpin2}
\widehat{\cal L}^{(2)}_{\pi N}&=& - \frac{1}{2M} \bar N_v \left[
D \cdot D +\frac{e}{2} (\mu_S+\tau_3\mu_V)
\varepsilon_{\mu \nu \rho \sigma} F^{\mu\nu} v^{\rho} S^{\sigma}\right]
N_v,\\
\label{lpin3}
\widehat{\cal L}^{(3)}_{\pi N}&=&
\frac{ie\varepsilon_{\mu\nu\rho\sigma} F^{\mu\nu}}{8 M^2} \bar N_v
\left[\mu_S-\frac{1}{2}+\tau_3(\mu_V-\frac{1}{2})\right]
S^{\rho} D^{\sigma} N_v +h.c.\,.
\end{eqnarray}

   The numerical results for the ten generalized proton polarizabilities 
are shown in Fig.\ \ref{fig:GPs} \cite{Hemmert_1997b}.
   As an example for $S=0$, let us discuss the generalized electric 
polarizability $\bar{\alpha}_E(\bar{q}^2)$, 
\begin{equation}
\label{alphaq2}
\frac{\bar{\alpha}_E(\bar{q}^2)}{\bar{\alpha}_E}=
1-\frac{7}{50}\frac{\bar{q}^2}{m^2_\pi}
+\frac{81}{2800}\frac{\bar{q}^4}{m^4_\pi}
+O\left(\frac{\bar{q}^6}{m^6_\pi}\right),\,\,
\bar{\alpha}_E=\frac{5 e^2 g_A^2}{384\pi^2m_\pi F_\pi^2}
=12.8\times 10^{-4}\,\mbox{fm}^3.
\end{equation}
   For $\bar{q}^2=0$, the generalized electric polarizability 
$\bar{\alpha}_E(\bar{q}^2)$ coincides with the electric polarizability 
$\bar{\alpha}_E$ of real Compton scattering 
\cite{Guichon_1995,Drechsel_1997a}. 
   The prediction of the $O(p^3)$ calculation for 
$\bar{\alpha}_E(0)=\bar{\alpha}_E$ agrees well with the
experimental value extracted from real Compton scattering,
$(12.1 \pm 0.8 \pm 0.5)\times 10^{-4}\,\mbox{fm}^3$ \cite{MacGibbon_1995}.
   In ChPT at $O(p^3)$ the generalized electric polarizability decreases
considerably faster with $\bar{q}^2$ than in the constituent quark model 
\cite{Guichon_1995}.
   Note that at $O(p^3)$, the results are entirely given in terms of the 
pion mass $m_\pi$, the axial coupling constant $g_A$, and the pion
decay constant $F_\pi$.
   This feature is true for all generalized polarizabilities.

   The $\pi^0$-exchange diagram only contributes to the spin-dependent GPs.
   As an example for a spin-dependent GP, let us consider
$P^{(11,11)1}$,
\begin{equation}
P^{(11,11)1}(\bar{q}^2)=-\frac{1}{288}\frac{g_A^2}{F_\pi^2}
\frac{1}{\pi^2 M}\left[\frac{\bar{q}^2}{m^2_\pi}-\frac{1}{10}
\frac{\bar{q}^4}{m^4_\pi}\right]
+\frac{1}{3M}\frac{g_A}{8\pi^2F^2_\pi}\tau_3
\left[\frac{\bar{q}^2}{m^2_\pi}-\frac{\bar{q}^4}{m^4_\pi}\right]
 +O\left(\frac{\bar{q}^6}{m^6_\pi}\right).
\end{equation}
   As a consequence of $C$ invariance, $P^{(11,11)1}(0)=0$ 
\cite{Drechsel_1997b} which, e.g., is not true for the
constituent quark model \cite{Guichon_1995}.
   The predictions for the spin-dependent GPs originate from two rather 
distinct sources---an isoscalar piece from pionic loop contributions,
and an isovector piece from the $\pi^0 \gamma\gamma^\ast$ vertex.
   It is interesting that the contributions of the pion-nucleon
loops to the spin-dependent GPs are much smaller
than the contributions arising from the $\pi^0$-exchange diagram.

\end{document}